\documentclass{ifacconf}

\usepackage{natbib}
\usepackage{graphicx}
\usepackage[utf8]{inputenc}
\usepackage{amsmath}
\usepackage{amssymb}
\usepackage{bm}
\usepackage[usenames,dvipsnames]{xcolor}
\usepackage{setspace}

\makeatletter
\newif\if@restonecol
\makeatother

\usepackage[lined,ruled]{algorithm2e}

\newcommand{\NiG}{\mathcal{N}i\varGamma}
\newcommand{\Nk}{\mathcal{N}_{k}}
\newcommand{\T}{^{\mathrm{T}}}

\def\argmin{\mathop{\arg\,\min}\limits}
\def\argmax{\mathop{\arg\,\max}\limits}
\begin{document}

\begin{frontmatter}

\title{Dynamic Bayesian Diffusion Estimation} 

\author[First]{Kamil Dedecius}
\author[First,Second]{, Vladim\'{i}ra Se\v{c}k\'{a}rov\'{a}}
\address[First]{\it Department of Adaptive Systems, Institute of Information Theory and Automation, Academy of Sciences of the Czech Republic, Prague, Czech Republic; (e-mail: dedecius@utia.cas.cz)}
\address[Second]{\it Department of Probability and Statistics, Faculty of Mathematics and Physics, Charles University in Prague, Czech Republic; (e-mail: seckarov@utia.cas.cz)}


\begin{abstract}
    The rapidly increasing complexity of (mainly wireless) ad-hoc networks stresses the need of reliable distributed estimation of several variables of interest. The widely used centralized approach, in which the network nodes communicate their data with a single specialized point, suffers from high communication overheads and represents a potentially dangerous concept with a single point of failure needing special treatment. This paper's aim is to contribute to another quite recent method called diffusion estimation. By decentralizing the operating environment, the network nodes communicate just within a close neighbourhood. We adopt the Bayesian framework to modelling and estimation, which, unlike the traditional approaches, abstracts from a particular model case. This leads to a very scalable and universal method, applicable to a wide class of different models. A particularly interesting case -- the Gaussian regressive model -- is derived as an example.
\end{abstract}

\end{frontmatter}

\section{Introduction}\label{S:Introduction}
We deal with the problem of collaborative estimation of unknown environmental parameter from noisy measurements. It naturally arises, e.g., in modern complex wireless systems and distributed sensor networks \citep{Aysal2008Constrained}. There exist two principal design schemes how to treat this estimation task: (i) the centralized approach, where the data are transmitted to a designated processing center (sometimes called fusion center) responsible for estimation (e.g., \citealp{Aysal2008Constrained} and many others); and (ii) the decentralized concept, where the nodes are responsible for estimation (e.g., \citet{xiao2006space,Cattivelli2010Diffusion}).
The decentralized methods become very promising, since the increasing complexity of modern networks calls for approaches with low overheads with respect to the time, energy and communication resources. Besides that, the potential single-points of failure (SPOFs) are principally avoided and a good design of the algorithm allows fast spatial reconfigurations of the network.

There exist several Bayesian methods treating general tasks with distributed character from the decision making perspective, ranging from \citep{Tsitsiklis1982Convergence} to \citep{Aysal2008Constrained}.
We focus ourselves on a recently formulated diffusion estimation problem, i.e., fully decentralized collaborative estimation in networks allowing the nodes to communicate only with their adjacent neighbours. In this field, a couple of non-Bayesian estimation algorithms were proposed. However, these are mostly single problem oriented, e.g., on least-squares estimation \citep{xiao2006space}, recursive least-squares (RLS, \citet{Cattivelli2008Diffusion}), least mean squares (LMS, \citet{Lopes2008Diffusion,Cattivelli2010Diffusion}), Kalman filters (\citet{Cattivelli2008Diffusion}) etc. We propose a new method called dynamic Bayesian diffusion estimation, which tackles the problem from the consistent and versatile Bayesian viewpoint and yields rather a methodology applicable to a much wider class of models, including, of course, the mentioned traditional ones. A particularly interesting application of the method to a Gaussian linear regressive model results in the so-called diffusion recursive least-squares method, proposed in \citet{Cattivelli2008Diffusion}. This demonstrates the generality of the method and advocates its feasibility. Furthermore, it shows that it is possible to shift from the viewpoint of a Bayesian statistician to the traditionalist's one, disregarding the probabilistic treatment of parameters of interest.

In this paper, we implicitly assume that the communication among nodes does not violate the bandwidth or other restrictions. The cases of restricted networks would require a specific solution which is behind the scope of this paper.

The organization of the paper is as follows: In Section 2, we briefly introduce the basic principle of Bayesian estimation. In Section 3, the dynamic Bayesian diffusion estimation theory is developed. Its application to the Gaussian linear regressive model follows in Section 4. Since we show that it leads to an existing solution, a demonstration example is avoided in the paper. We conclude our work and outline the future research topics in Section \ref{S:conclusion}.

\section{Bayesian estimation}\label{S:distributed}
Let us consider a linear stochastic system with a real input variable $u_{t}$ and a real output variable $y_t$, observed at discrete time instants $t=1,2,\dots$ Both $u_{t}$ and $y_{t}$ can be scalar or multivariate. We form a data $\bm{d}(t)$ as an ordered set of observations and inputs, $\bm{d}(t) = \{y_0, u_0, \dots, y_t, u_t\}$. The dependence of the output $y_t$ on the previous data ${\bm{d}(t-1)}$ and the current input $u_{t}$ can be modelled by a conditional probability density function (pdf)
\begin{equation}
    f(y_t|u_t, \bm{d}(t-1), \bm{\Theta}),
    \label{E:model}
\end{equation}
where $\bm{\Theta}$ is a random potentially multivariate model parameter.

The Bayesian methodology treats the model parameter as an unobservable random variable whose knowledge at time $t$ is carried by past data $\bm{d}(t-1)$. The Bayesian estimation of $\bm{\Theta}$ then exploits pdf $g(\bm{\Theta}|\bm{d}(t-1))$. By the assumption of natural conditions of control \citep{peterka1981bayesian} we have
\begin{equation}
    g(\bm{\Theta}|u_t,\bm{d}(t-1)) = g(\bm{\Theta}|\bm{d}(t-1)),
    \label{E:nnc}
\end{equation}
i.e., the information about parameter $\bm{\Theta}$ at time $t$ is conditionally independent of the current input $u_t$.
 The prior knowledge $\bm{d}(0) = \{y_{0}, u_{0}\}$ formed by the initial data can be determined by an expert or it follows from past estimation. It is also possible to start from a noninformative (flat) prior pdf.

The Bayesian recursive estimation exploits the Bayes' rule to incorporate new data into the prior pdf of $\bm{\Theta}$ as follows
\begin{equation}
    g(\bm{\Theta}|\bm{d}(t)) \propto
    f(y_t|u_{t},\bm{d}(t-1), \bm{\Theta})
    g(\bm{\Theta}|\bm{d}(t-1)),
    \label{E:update}
\end{equation}
where $\propto$ denotes equality up to a normalizing constant. At the next time instant, the posterior pdf on the left-hand side of \eqref{E:update} is used as the prior pdf. The last relation is also known as the dynamic Bayesian data update.

\section{Dynamic Bayesian diffusion estimation}\label{S:diffusion}
Let us now focus on the diffusion estimation task. Let there be a distributed network consisting of a set of nodes interacting with their neighbours, which collectively estimate the common parameter of interest using the same model structure. Furthermore, let us impose the following constraint: \emph{the nodes are able to communicate one-to-one only within their closed neighbourhood} defined as follows:
\begin{defn}
    Given a network represented by an undirected graph consisting of $M \in \mathbb{N}$ nodes, the closed neighbourhood $\Nk$ of the $k$th node, $1 \leq k \leq M$, is the set consisting of its adjacent nodes and node $k$.
\end{defn}

\begin{figure}[htb]
    \begin{center}
        \includegraphics[scale=.4]{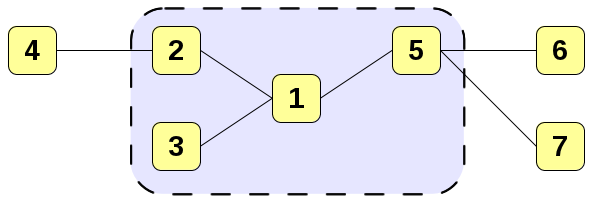}
    \end{center}
    \caption{Closed neighbourhood $\mathcal{N}_{1}=\left\{ 1,2,3,5 \right\}$.}
    \label{F:net}
\end{figure}

An example of a network including a closed neighbourhood $\mathcal{N}_{1} = \left\{ 1,2,3,5 \right\}$ of node $k=1$ is depicted in Figure \ref{F:net}.

The diffusion estimation involves two subsequent steps, the former of which is optional but preferred:
\begin{description}
    \item [Incremental update] -- also known as the data update, is a diffusion alternative of \eqref{E:update}. The nodes propagate data within their closed neighbourhood and incorporate them into their local statistical knowledge;
    \item [Spatial update] -- the nodes propagate point parameter estimates (i.e. mean values) or posterior pdfs within their closed neighbourhood and correct their local estimates.
\end{description}

\begin{figure}[htb]
    \begin{center}
        \includegraphics[scale=.5]{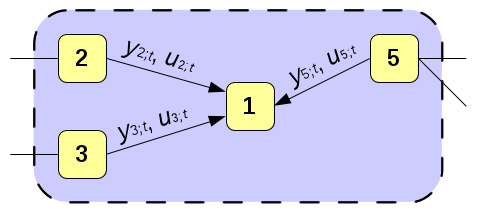}
    \end{center}
    \caption{Incremental update of node $k=1$ by data from its adjacent neighbours $l\in\Nk$. The spatial update looks similarly, the nodes exchange either whole pdfs (i.e., the hyperparameters) of $\bm{\Theta}$ or its estimates.}
    \label{F:incremental}
\end{figure}

\subsection{Incremental update}
First, we develop the general theory of the incremental update using the Bayesian decision making paradigm.
Let $A$ be a measurable space of decisions, $\beta = \dim(\bm{\Theta})$  and let $L: \mathbb{R}^{\beta} \times A \to \mathbb{R}$ be an $\mathbb{L}_{1}$-measurable loss function. The Bayesian decision making problem consists of choosing $a \in A$ by using a measurable decision rule $\delta: \mathbb{R} \to A$ after an observation of random variable $X$ being obtained. Therefore, we introduce the risk
\begin{equation}
    R\left( \bm{\Theta}, \delta \right)
    = \mathbb{E}_{X} \left[ L(\bm{\Theta}, \delta(X))|\bm{\Theta}\right]
    \label{E:risk}
\end{equation}
and the Bayesian risk function
\begin{equation}
    \rho(g,\delta)=\mathbb{E}_{\bm{\Theta}}\left[R\left( \bm{\Theta}, \delta \right)\right]
    \label{E:risk2}
\end{equation}
measuring the quality of a decision rule $\delta$ under ignorance of a parameter $\bm{\Theta}$ with prior $g(\bm{\Theta})$. The Bayes' rule is that one which satisfies the condition
\begin{equation}
    \mathbb{E}_{\bm{\Theta}} \left[ L(\bm{\Theta}, \delta(X)) | X=x \right]
    = \inf_{a \in A} \mathbb{E}_{\bm{\Theta}}\left[ L(\bm{\Theta}, a)|X=x \right]
    \label{E:minrisk}
\end{equation}
where the integration is with respect to the posterior pdf of $\bm{\Theta}$.

Consider now the situation from the $k$th node's perspective, exploiting the data from its closed neighbourhood.
In \citep{stone1977consistent}, for any given $a$ and weights $c_{l,k}$ (where $l\in\Nk$), the approximate of the Bayesian inference under ignorance of the prior distribution was proposed in terms of
\begin{equation}
    \widehat{\mathbb{E}} \left[ L_{k}(\bm{\Theta}, a) | X=x \right]
    = \sum_{l\in\Nk} c_{l,k} L_{l}(\bm{\Theta}, a).
    \label{E:stone}
\end{equation}
Namely, $c_{l,k}$ represents weight of $l$th node with respect to the $k$th one and $\sum_{l\in\Nk}c_{l,k} = 1$.

Remind, that the Bayes' rule transforming the prior pdf to the posterior pdf is completely compatible with the maximum entropy principle \citep{Giffin2007Updating}, hence we only need to reflect the fact that for a fixed time, multiple data are at disposal. To stay in the entropy framework, we will exploit the minimum cross-entropy principle (MinXEnt) to find a rule for handling the data.
\begin{defn}[Kullback Leibler divergence] {\ \\}
    Let $f$, $g$ be two pdfs describing random variable $X$. The Kullback-Leibler divergence (also known as the cross-entropy) of $f$ and $g$ is defined as
    \begin{align}
        \mathcal{D}(f||g) &= \int f(x) \log \frac{f(x)}{g(x)} \mathrm{d}x \notag \\
                &= \int f(x) \log f(x) \mathrm{d}x - \int f(x) \log g(x) \mathrm{d}x \notag \\
                &= H(f,g) - H(f) \label{E:KL}
    \end{align}
    where $H(\cdot)$ denotes entropy and $H(\cdot,\cdot)$ stands for the cross-entropy.
\end{defn}
\begin{cor}
    Given $f$, the minimization of the Kullback-Leibler divergence $\mathcal{D}(f||g)$ is equivalent to the minimization of $H(f,g)$.
\end{cor}
\begin{pf}
    Trivial.
\end{pf}

Instead of operating on nodes' posterior pdfs using a sort of averaging or projection, e.g. \citep{KNiha}, we propose to exploit the principle of weighted likelihoods \citep{Wang2004Asymptotic,Wang2006Approximating}.
Let $f(x|\bm{\Theta})$ and $f(x|a)$ denote conditional pdfs with respect to $\bm{\Theta}$ and $a$ respectively. The Bayesian framework assigns $\mathcal{D}(f(x|\bm{\Theta})||f(x|a)) = L(\bm{\Theta}, a)$.
Under $k$ fixed, \eqref{E:stone} reads
\begin{equation*}
    \widehat{\mathbb{E}} \left[ L_{k}(\bm{\Theta}, a)|x_{k} \right]
    = \sum_{l\in \Nk} c_{l,k} \mathcal{D}\!\left(f_{l}(x_l|\bm{\Theta})\big|\big|f_{l}(x_l|a)\right),
\end{equation*}
where $x_{l}$ denotes data from $l$th node. Since we have just one observation for each node $l\in\Nk$, we get
\begin{equation}\label{E:estrisk}
    \widehat{\mathbb{E}} \left[ L_{k}(\bm{\Theta}, a)|x_{k} \right]
    = \sum_{l\in \Nk} c_{l,k} f_{l}(x_l|\bm{\Theta})\log\frac{f_{l}(x_l|\bm{\Theta})}{f_{l}(x_l|a)}.
\end{equation}
Under ignorance of $\bm{\Theta}$ we set, accordingly to maximum entropy principle, $f_{l}(x_l|\bm{\Theta})=1/\textrm{card}(\Nk)$ where $\textrm{card}$ denotes set cardinality. Formula \eqref{E:estrisk} then looks as follows:
\begin{equation}\label{E:resultrisk}
    \sum_{l\in \Nk} \frac{c_{l,k}}{\textrm{card}(\Nk)}\log f_{l}(x_l|\bm{\Theta})-\sum_{l\in \Nk} \frac{c_{l,k}}{\textrm{card}(\Nk)}\log f_{l}(x_l|a).
\end{equation}

We see that only the second part of \eqref{E:resultrisk} should be considered for the minimization "through" the set $A$ of possible decisions. Particularly:
\begin{gather}
    \argmin_{a\in A} \left( -\sum_{l\in\Nk} c_{l,k}\log f_{l}(x_l|a) \right) \notag\\[2mm]
    = \argmax_{a\in A} \sum_{l\in\Nk} c_{l,k}\log f_{l}(x_l|a) \notag \\[2mm]
    = \argmax_{a\in A} \prod_{l\in\Nk} f_{l}(x_l|a)^{c_{l,k}},
    \label{E:argmin}
\end{gather}
where $c_{l,k}$ denote the previously mentioned weights.
The argument \eqref{E:argmin} together with the Bayes' rule \eqref{E:update}, preserving entropy maximization, yield theoretically consistent incremental update in the form
\begin{align}
    g_{k}(\bm{\Theta}|\overline{\bm{d}}(t)) &\propto g_{k}(\bm{\Theta}|\overline{\bm{d}}(t-1)) \notag \\[2mm]
    &\times \prod_{l\in\Nk} f_{l}(y_{l,t}|u_{l,t}, \bm{d}_{l}(t-1), \bm{\Theta})^{c_{l,k}},
    \label{E:diff-update}
\end{align}
where $\overline{\bm{d}}(t)$ stands for all data available from sources in $\Nk$.

\subsection{Spatial update}
The spatial update follows after the incremental update. In this step, the nodes exchange information about unknown model parameter $\bm{\Theta}$, either in the form of its estimates or hyperparameters of its distribution.
Formally, for fixed $k$, the information from all nodes in $\Nk$ describes the finite mixture density
\begin{equation}
    g_{k}(\bm{\Theta}|\overline{\bm{d}}(t)) = \sum_{l\in\Nk} a_{l,k} g_l(\bm{\Theta}|\overline{\bm{d}}(t)), \quad \sum_{l\in\Nk}a_{l,k} = 1,
    \label{E:spatial-update-mix}
\end{equation}
where $0\leq a_{l,k}\leq 1$ is the weight of $l$th node's estimate from $k$th node's viewpoint.

Here, two possible departure points arise. First, more generally, we may be interested in a ``consensus'' distribution, i.e., a single distribution best representing the mixture \eqref{E:spatial-update-mix} at node $k$. Its pdf can be found as the argument minimizing the Kullback-Leibler divergence,
\begin{equation}
    \argmin_{\tilde{g}_{k}(\bm{\Theta}|\overline{\bm{d}}(t))\in\mathcal{G}} \mathcal{D}
    \left(
    g_{k}(\bm{\Theta}|\overline{\bm{d}}(t))\bigg|\bigg| \tilde{g}_{k}(\bm{\Theta}|\overline{\bm{d}}(t))
    \right),
    \label{E:spatial-update}
\end{equation}
where $\mathcal{G}$ is the class of all admissible pdfs.

The second possibility emerges if we are interested just in the moment(s) available from $g_{k}(\bm{\Theta}|\overline{\bm{d}}(t))$. Then, e.g., the first moment (the mean value) is given by the convex combination of mean values of the mixture density components,
\begin{equation}
    \widehat{\bm{\Theta}}_{k} \leftarrow \sum_{l\in\Nk} a_{l,k} \widehat{\bm{\Theta}}_l.
    \label{E:spatial-update-moments}
\end{equation}
For other moments see, e.g., \citet{FruhwirthSchnatter2006Finite}. The latter approach is of particular interest if the distribution is parameterized by moments (e.g., the Gaussian distribution). Another appealing fact related to these distributions is that \eqref{E:spatial-update-moments} is often a direct consequence of \eqref{E:spatial-update}. In these cases, it is possible to omit the Kullback-Leibler divergence minimization and benefit directly from \eqref{E:spatial-update-moments}.
While \eqref{E:spatial-update-moments} is a final product at time $t$, the pdf resulting from \eqref{E:spatial-update} can be reused as the $k$'s prior pdf at the next time step.

Properties of the diffusion estimator strongly depend on the underlaying particular estimators in a neighbourhood and their weights $a_{l,k}$ and $c_{l,k}$. In this respect, the need for effective determination of weights is essential.

\subsection{Determination of weights $a_{l,k}$ and $c_{l,k}$}
There are several possible strategies how to determine the weights $a_{l,k}$ and $c_{l,k}$. Besides the relatively unfeasible uniform weights, the user can perform with the aid of Metropolis weights, proposed by \citet{xiao2006space} and further used in recent literature. Another options are relative degree and yet more sophisticated relative degree-variance weights, based on the cardinality of the node's closed neighbourhood, \citep{Cattivelli2010Diffusion}. We only conjecture that a suitable probabilistic method exploiting, e.g., the likelihood of $l$th data with respect to $k$th node could be found as well. A substantial advantage of such method would be its suitability for dynamic cases, requiring stable determination of $a_{l,k}$ and $c_{l,k}$. As a consequence, it would allow to suppress the influence of data and/or estimates from a failing node (sensor) on other nodes. However, such methods are being developed in the meantime.

\section{Derivation for Gaussian regressive model}\label{S:derivation}
In this section, a practical application of the proposed methodology is given. We derive the dynamic Bayesian diffusion estimator of the popular Gaussian linear regressive model. In two following subsections, we shortly present the standard Bayesian estimation of such model and develop its diffusion estimator.
This case is just one example of a wide class of possible models, the applicability on which is straightforward. This class includes particularly popular Bayesian models with conjugate priors.

\subsection{Gaussian linear regressive model}
Given a regression vector $\bm{\psi}_t \in \mathbb{R}^{n}, t=1,2,\dots$ and a dependent random variable $y_t \in \mathbb{R}$, the Gaussian linear regressive model takes the form
\begin{equation}
    y_t = \bm{\psi}_{t}\T \bm{\theta} + \varepsilon_{t},
    \label{E:linear-regression}
\end{equation}
where $\bm{\theta}\in\mathbb{R}^{n}$ is the regression coefficient and ${\varepsilon_{t} \sim \mathcal{N}(0,\! \sigma^2)}$ is the Gaussian white noise. This makes ${y_t \sim \mathcal{N}(\bm{\psi}_t\T \bm{\theta}, \sigma^{2})}$ and the regression model \eqref{E:linear-regression} can be expressed by pdf ${f(y_{t}|\bm{\psi}_{t}, \bm{\Theta}})$. From the Bayesian viewpoint, the model parameters ${\bm{\Theta} \equiv \{\bm{\theta}, \sigma^{2}\}}$ are also random variables.
Under ignorance of their values, the proper conjugate prior distribution is the normal inverse-gamma ($\NiG$) one \citep{Bernardo1994Bayesian}. Namely, $\bm{\theta}$ is normal and $\sigma^{2}$ is inverse-gamma.

\begin{defn}[Normal inverse-gamma pdf]{\ \\}
    For a variable $\bm{\Theta} = \{\bm{\theta}, \sigma^{2}\}$,  $\bm{\theta} \in \mathbb{R}^{n}$ and ${\sigma^{2} \in \mathbb{R}}$, the normal inverse-gamma $\NiG(\bm{V}, \nu)$ pdf with a symmetric positive definite extended information matrix ${\bm{V}\in \mathbb{R}^{N\times N}}, {N=n+1}$ and the degrees of freedom $\nu\in\mathbb{R}$ has the form
    \begin{equation}
        g(\bm{\theta}, \sigma^2|\bm{V},\nu)
        = \frac{\sigma^{-(\nu+n+1)}}{\mathcal{I}(\bm{V}, \nu)}
        \!\exp\!
        \left\{\!
        -\frac{1}{2\sigma^{2}}
        \begin{bmatrix}
            -1 \\
            \bm{\theta}
        \end{bmatrix}\T
        \!\bm{V}
        \begin{bmatrix}
            -1 \\
            \bm{\theta}
        \end{bmatrix}
        \right\}
        \notag
    \end{equation}
where $\mathcal{I}(\cdot)$ is the normalization term such that
\begin{equation}
    \int g(\bm{\theta}, \sigma^2|\bm{V},\nu) \mathrm{d}\bm{\Theta} = 1.\notag
\end{equation}
\end{defn}
Both $\bm{V}$ and $\nu$ are sufficient statistics \citep{Bernardo1994Bayesian} representing data $\bm{d}(t-1) = \{y_{t-1}, \psi_{t-1}, \dots, y_{0}, \psi_{0}\}$. The Bayesian recursive estimation \eqref{E:update} updates the prior pdf by new data according to the following theorem.

\begin{thm}[Bayesian estimation of a $\NiG$ model]\label{T:bayesian-estimation-nig}{\ \\}
    Let $g(\bm{\theta}, \sigma^{2}|\bm{V}, \nu)$ be a $\NiG$ pdf, $t=1,2,\dots$ The Bayesian estimation \eqref{E:update} updates the sufficient statistics $\bm{V}\in\mathbb{R}^{N\times N}$ and $\nu \in \mathbb{R}$  by real scalar realization $y_{t}$ and regression vector $\bm{\psi}_{t} \in \mathbb{R}^{N-1}$ as follows:
    \begin{align}
        \bm{V}_{t} &= \bm{V}_{t-1}
        +
        \begin{bmatrix}
            y_{t} \\
            \bm{\psi}_{t}
        \end{bmatrix}
        \begin{bmatrix}
            y_{t} \\
            \bm{\psi}_{t}
        \end{bmatrix}\T \label{E:update-v} \\
        \nu_{t} &= \nu_{t-1} + 1 \label{E:update-nu}
    \end{align}
    The multivariate point estimator $\widehat{\bm{\theta}}_t \in \mathbb{R}^{N-1}$ of regression coefficient is the mean value of the $\NiG$ distribution given by
    \begin{equation}
        \widehat{\bm{\theta}}_{t}
        =
        \begin{bmatrix}
            V_{22} & \ldots & V_{2N} \\
            \vdots & \ddots & \vdots \\
            V_{N2} & \ldots & V_{NN}
        \end{bmatrix}^{-1}
        _{t}
        \begin{bmatrix}
            V_{21} \\ \vdots \\ V_{N1}
        \end{bmatrix}_{t}
        \label{E:point-est}
    \end{equation}
\end{thm}
\begin{pf}
    The update of statistics $\bm{V}$ and $\nu$ follows directly from multiplication of Gaussian models (likelihoods), see, e.g., \cite{peterka1981bayesian}. The point estimator is the well-known ordinary least squares estimator.
\end{pf}

\subsection{Diffusion estimation of the Bayesian regressive model}
In order to derive the dynamic Bayesian diffusion estimator of $\bm{\Theta}$, we follow the principles given in Section \ref{S:diffusion}.
Let us consider a network of $M\in\mathbb{N}$ distributed nodes. Each node ${k\in\{1,\dots,M\}}$  evaluates a model
\begin{equation}
    f(y_{k;t}|\bm{\psi}_{k;t}, \bm{\Theta}, \bm{V}_{k;t-1}, \nu_{k;t-1})
    \label{E:gauss-models}
\end{equation}
and runs the diffusion Bayesian estimation \eqref{E:diff-update} of its parameters in the form
\begin{align}
    g_{k}(\bm{\Theta}|\bm{V}_{k;t}, \nu_{k;t})&\propto
    g_{k}(\bm{\Theta}|\bm{V}_{k;t-1}, \nu_{k;t-1})  \notag \\[2mm]
    &\!\!\! \!\!\!\times \prod_{l\in\Nk} f_{l}(y_{l;t}|\bm{\psi}_{l;t}, \bm{\Theta}, \bm{V}_{l;t-1}, \nu_{l;t-1})^{c_{l,k}}. \label{E:diff-gauss}
\end{align}
Here $0 \leq c_{l,k} \leq 1$ weights $l$th node's data with respect to $k$th node, $l\in\Nk$, where $\sum_{l\in\Nk} c_{l,k} = 1$.
Simply put, the $k$th node updates its prior pdf of $\bm{\Theta}$ by data from its closed neighbourhood $\Nk$. Since we deal with the $\NiG$ pdf, this update takes the form expressed by the following proposition.

\begin{prop}[Incremental update of $\NiG$ pdf]\label{P:gauss-incremental-update}
    Given a $k$th node, $k\in\{1,\dots,M\}$, the incremental version of the Bayesian estimation (Theorem \ref{T:bayesian-estimation-nig})  updates the $k$th node's prior $\NiG$ pdf of $\bm{\Theta}$ by data $[y_{l;t}, \bm{\psi}_{l,t}]\T$, weighted by $c_{l,k}$, from its adjacent neighbours $l\in\Nk$ according to the following rules:
\begin{align}
    \bm{V}_{k;t} &= \bm{V}_{k;t-1}
    +
    \sum_{l \in \Nk}
    c_{l,k}
    \begin{bmatrix}
        y_{l;t} \\
        \bm{\psi}_{l;t}
    \end{bmatrix}
    \begin{bmatrix}
        y_{l;t} \\
        \bm{\psi}_{l;t}
    \end{bmatrix}\T \label{E:diff-update-v} \\
    \nu_{k;t} &= \nu_{k;t-1} + 1, \label{E:diff-update-nu}
\end{align}
where
\begin{equation}
    0 \leq c_{l,k} \leq 1, \qquad \sum_{l\in\Nk} c_{l,k} = 1, \quad l\in\Nk.
    \notag
\end{equation}
\end{prop}

\begin{pf}
    Let $\kappa = \mathrm{card}(\Nk)$. The formula \eqref{E:diff-update-v} following from \eqref{E:diff-gauss} is equivalent to $\kappa$ updates \eqref{E:update-v} of $\bm{V}_{k,t-1}$  by data $[y_{l,t}, \bm{\psi}_{l,t}]\T$ weighted by $c_{l,k}$. Formula \eqref{E:diff-update-nu} is a direct equivalent of \eqref{E:update-nu}. \phantom{xxxxxxxxxxxxxxxxxxxxxxxxxxxxxxxxx} \qed
\end{pf}

In linear regression, we are particularly interested in point estimation of the regression coefficient $\bm{\theta}$. 

\begin{prop}[Spatial update of $\widehat{\bm{\theta}}$]\label{P:gauss-spatial-update}
    Given a $k$th node, $k\in\{1,\dots,M\}$. The spatial update \eqref{E:spatial-update-moments} of the estimate $\hat{\bm{\theta}}_{k;t}$ has the form
\begin{equation}
    \hat{\bm{\theta}}_{k;t} = \sum_{l\in\Nk} a_{l,k} \hat{\bm{\theta}}_{l;t},
    \label{E:diff-update-spatial}
\end{equation}
where
\begin{equation}
    0 \leq a_{l,k} \leq 1, \qquad \sum_{l\in\Nk} a_{l,k} = 1.\notag
\end{equation}
$a_{l,k}$ denotes the weight of $l$th node's point estimate with respect to $k$th node.
\end{prop}
\begin{pf}
    This is a straightforward use of \eqref{E:spatial-update-moments}.\phantom{xxxxxxx}  \qed
\end{pf}
Similar procedure applies to estimation of $\sigma^{2}$.
The summary of the derived steps is in Algorithm 1.

\section{Dynamic Bayesian diffusion regressive model and RLS}
Let us demonstrate the simplicity of transition from the dynamic Bayesian diffusion estimation to its non-Bayesian counterpart. For simplicity, consider $y$ scalar and partition the extended information matrix $\bm{V}$ as follows:
\begin{equation}
    \bm{V} =
    \left[
    \begin{array}{c|c}
        V_{y} & \bm{V}_{y\psi}\T \\ \hline
        \bm{V}_{y\psi} & \bm{V}_{\psi}
    \end{array}
    \right],
    \text{ where }
    V_{y} \in \mathbb{R},  \bm{V}_{\psi} \in \mathbb{R}^{n\times n}.
    \label{E:partitioned-v}
\end{equation}
Furthermore, let us denote $\bm{C} = \bm{V}_{\psi}^{-1}$ and see, how the update -- Proposition \ref{P:gauss-incremental-update} -- performs on reparameterized $\NiG$ pdf.
\begin{prop}[Reparametrization of $\NiG$ pdf]{\ \\}
    Given pdf $\NiG(\bm{V},\nu)$ of $\bm{\Theta} = \{\bm{\theta}, \sigma^{2}\}$. The statistic $\bm{V}\in\mathbb{R}^{N\times N}$ can be decomposed into the lower-dimensional statistics $\bm{C}\in \mathbb{R}^{n\times n}, \widehat{\bm{\theta}} \in \mathbb{R}^{n}$ and $\Lambda \in \mathbb{R}$ where $n = N-1$, yielding the reparametrized pdf $\NiG(\bm{C}, \widehat{\bm{\theta}}, \Lambda, \nu)$ as follows:
    \begin{align}
        g&(\bm{\theta}, \sigma^2|\bm{C},\widehat{\bm{\theta}}, \Lambda, \nu)
        = \frac{\sigma^{-(\nu+n+1)}}{\mathcal{I}(\bm{C},\widehat{\bm{\theta}}, \Lambda, \nu)} \times \notag \\
        &\times
        \exp
        \left\{
            -\frac{1}{2\sigma^{2}}
            \left[
                (\bm{\theta} - \widehat{\bm{\theta}})\T
                \bm{C}^{-1}
                (\bm{\theta} - \widehat{\bm{\theta}})
                + \Lambda
            \right]
        \right\}
    \end{align}
    where
    \begin{align}
        \widehat{\bm{\theta}} &= \bm{C} \bm{V}_{y\psi} \label{E:hat-theta}, \\
        \Lambda &= V_{y} - \bm{V}_{y\psi}\T \bm{C} \bm{V}_{y\psi} \label{E:lambda}
    \end{align}
    and where $\mathcal{I}(\bm{C},\widehat{\bm{\theta}}, \Lambda, \nu)$ is the normalization term such that
    \begin{equation}
        \int g(\bm{\theta}, \sigma^2|\bm{C},\widehat{\bm{\theta}}, \Lambda, \nu) \mathrm{d}\bm{\Theta} = 1. \notag
    \end{equation}
\end{prop}

\begin{pf}
    By completion of squares
    \begin{gather*}
            \begin{bmatrix}
            -1 \\
            \bm\theta
            \end{bmatrix}\T
            \left[
            \begin{array}{cc}
                V_{y} & \bm{V}_{y\psi}\T \\
                \bm{V}_{y\psi} & \bm{V}_{\psi}
            \end{array}
            \right]
            \begin{bmatrix}
            -1 \\
            \bm\theta
            \end{bmatrix}
        =
        V_{y} - 2 \bm{\theta}\T \bm{V}_{y\psi} + \bm{\theta}\T \bm{V}_{\psi} \bm{\theta}
        \\[2mm]
        =
        \left( \bm{\theta} - \bm{C} \bm{V}_{y\bm{\psi}} \right)^{\!\mathrm{T}}
        \!\!\bm{C}^{-1}\!
        \left( \bm{\theta} - \bm{C} \bm{V}_{y\bm{\psi}} \right)
        \!+\!
        \left( V_{y} - \bm{V}_{y\psi}\T \bm{C} \bm{V}_{y\psi} \right)
        .
    \end{gather*}
    \flushright\qed
\end{pf}
Now, we focus on the recursive update of $k$th node's reparameterized $\NiG$ pdf statistics. First note, that the right-hand side of formula \eqref{E:diff-update-v} can be viewed as a sequential (one-by-one) update of $k$th nodes' $\bm{V}_{k,t}$ by data $[y_{l;t}, \bm{\psi}_{l,t}]\T$ with weights $c_{l,k}$ where $l\in\Nk$. This means, that when the transition $(t-1) \to t$ occurs, the assignment
\begin{equation}
    \bm{V}_{k;t} := \bm{V}_{k;t-1}
    \label{E:V-assign}
\end{equation}
is made, followed by the updates
\begin{equation}
    \bm{V}_{k;t}
    \leftarrow \bm{V}_{k;t} + c_{l,k}
    \begin{bmatrix}
        y_{l;t} \\
        \bm{\psi}_{l,t}
    \end{bmatrix}
    \begin{bmatrix}
        y_{l;t} \\
        \bm{\psi}_{l;t}
    \end{bmatrix}\T
    \quad \text{for all}\quad l\in\Nk.
    \label{E:V-update1}
\end{equation}
Therefore, we can take advantage of deriving the update of $k$th reparameterized pdf by data from $l$th node. The reparameterized equivalent of \eqref{E:diff-update-v} then results from \eqref{E:V-update1} for all $l\in\Nk$ and $t$ fixed. This sequential update procedure describes the following proposition.

\begin{prop}[Update of reparameterized $\NiG$ pdf] {\ \\}
    Given a pdf $g(\bm{\theta}, \sigma^{2}|\bm{C}, \widehat{\bm{\theta}}, \Lambda, \nu)$ of $k$th node at fixed time~$t$. After initialization
    \begin{align}
        \bm{C}_{k,t}&:=\bm{C}_{k,t-1},&\widehat{\bm{\theta}}_{k;t}&:=\widehat{\bm{\theta}}_{k;t-1}, \notag \\
        \Lambda_{k;t}&:= \Lambda_{k;t-1},&\nu_{k;t}&:=\nu_{k;t-1}, \label{E:init}
    \end{align}
the update by data $y_{l;t}, \bm{\psi}_{l;t}$, weighted by $c_{l,k}$ for all $l\in\Nk$ reads
{\allowdisplaybreaks
\begin{align}
    \bm{C}_{k;t} &\leftarrow \bm{C}_{k;t}
        - \frac{c_{l,k} \bm{C}_{k;t} \bm{\psi}_{l;t}\bm{\psi}_{l;t}\T \bm{C}_{k;t}}
               {1 + c_{l,k} \bm{\psi}_{l;t}\T \bm{C}_{k;t} \bm{\psi}_{l;t}} \label{E:rls-C} \\
    \widehat{\bm{\theta}}_{k;t} &\leftarrow \widehat{\bm{\theta}}_{k;t}
        +\frac{c_{l,k} \bm{C}_{k;t} \bm{\psi}_{l;t}}
              {1 + c_{l,k} \bm{\psi}_{l;t}\T \bm{C}_{k;t} \bm{\psi}_{l;t}}
        [y_{l;t}-\bm{\psi}_{l;t}\T\widehat{\bm{\theta}}_{k;t}] \label{E:rls-theta} \\
    \Lambda_{k;t} &\leftarrow \Lambda_{k;t}+
    \frac{\left(c_{l,k}y_{l;t}+c_{l,k}\bm{\psi}_{k;t}\T\widehat{\bm{\theta}}_{k;t}\right)^2}
             {1 + c_{l,k} \bm{\psi}_{l;t}\T \bm{C}_{k;t} \bm{\psi}_{l;t}} \label{E:rls-lambda} \\
             \nu_{k;t} &\leftarrow \nu_{k;t} + c_{l,k} \label{E:rls-nu}
\end{align}
}
\end{prop}

\begin{pf}
    Fix $t$ and rewrite the update of blocks of $\bm{V}_{k;t}$ of $k$th node by $y_{l;t}$ and $\bm{\psi}_{l;t}$ from its adjacent neighbour $l\in\Nk$. The initialization \eqref{E:init} is equivalent to
\begin{equation}
    \bm{V}_{k;t} \leftarrow \bm{V}_{k;t-1}, \qquad
    \nu_{k;t} \leftarrow \nu_{k;t-1}. \notag
\end{equation}

The blocks of $\bm{V}_{k;t}$ are updated as follows:
\begin{align}
    V_{k;y;t} &\leftarrow V_{k;y;t} + c_{l,k} y_{l;t}^{2} \label{E:rls-Vy} \\
    \bm{V}_{k;\psi;t} &\leftarrow \bm{V}_{k;\psi;t} +  c_{l,k} \bm{\psi}_{l;t} \bm{\psi}_{l;t}\T \label{E:rls-Vpsi} \\
    \bm{V}_{k;y\psi;t} &\leftarrow \bm{V}_{k;y\psi;t} + c_{l,k} \bm{\psi}_{l;t} y_{l;t} \label{E:rls-Vypsi}
\end{align}

Notice, that \eqref{E:rls-Vpsi} is equivalent to
\begin{equation}
    \bm{C}_{k;t}^{-1} \leftarrow \bm{C}_{k;t}^{-1} + c_{l,k} \bm{\psi}_{l;t} \bm{\psi}_{l;t}\T.
\end{equation}

By application of the Sherman-Morrison formula, Proposition \ref{A:sherman-morrison} in Appendix, we obtain
\begin{equation}
    \bm{C}_{k;t} \leftarrow \bm{C}_{k;t}
    - \frac{c_{l,k} \bm{C}_{k;t} \bm{\psi}_{l;t}\bm{\psi}_{l;t}\T \bm{C}_{k;t}}
           {1 + c_{l,k} \bm{\psi}_{l;t}\T \bm{C}_{k;t} \bm{\psi}_{l;t}}, \notag
\end{equation}
which proves \eqref{E:rls-C}.

The substitution of \eqref{E:rls-C} and \eqref{E:rls-Vypsi} into \eqref{E:hat-theta} yields
{\allowdisplaybreaks
\begin{align}
    \widehat{\bm{\theta}}_{k;t} 
    &\leftarrow \left(\!\!\bm{C}_{k;t}
    - \frac{c_{l,k}\! \bm{C}_{k;t} \bm{\psi}_{l;t}\bm{\psi}_{l;t}\T \bm{C}_{k;t}}
           {1 + c_{l,k} \bm{\psi}_{l;t}\T \bm{C}_{k;t} \bm{\psi}_{l;t}}\right)
    \!\!\left(\bm{V}_{k;y\psi;t} + c_{l,k} \bm{\psi}_{l;t} y_{l;t}\right) \notag \\[1mm]
    &\leftarrow {\bm{C}_{k;t}\bm{V}_{k;y\psi;t}} + c_{l,k}\bm{C}_{k;t}\bm{\psi}_{l;t} y_{l;t}
    -\frac{c_{l,k} \bm{C}_{k;t} \bm{\psi}_{l;t}\bm{\psi}_{l;t}\T \bm{C}_{k;t}}
          {1 + c_{l,k} \bm{\psi}_{l;t}\T \bm{C}_{k;t} \bm{\psi}_{l;t}} \notag \\[1mm]
    &\times \bm{V}_{k;y\psi;t}
    - \frac{c_{l,k} \bm{C}_{k;t} \bm{\psi}_{l;t}\bm{\psi}_{l;t}\T \bm{C}_{k;t}}{1 + c_{l,k} \bm{\psi}_{l;t}\T \bm{C}_{k;t} \bm{\psi}_{l;t}}c_{l,k} \bm{\psi}_{l;t} y_{l;t} \notag \\[1mm]
    {} & \leftarrow \widehat{\bm{\theta}}_{k;t}+\frac{c_{l,k} \bm{C}_{k;t} \bm{\psi}_{l;t}}{1 + c_{l,k} \bm{\psi}_{l;t}\T \bm{C}_{k;t} \bm{\psi}_{l;t}}[y_{l;t}-\bm{\psi}_{l;t}\T{\bm{C}_{k;t}\bm{V}_{k;y\psi;t}}]\notag \\[1mm]
    {} & \leftarrow \widehat{\bm{\theta}}_{k;t}+\frac{c_{l,k} \bm{C}_{k;t} \bm{\psi}_{l;t}}{1 + c_{l,k} \bm{\psi}_{l;t}\T \bm{C}_{k;t} \bm{\psi}_{l;t}}[y_{l;t}-\bm{\psi}_{l;t}\T\widehat{\bm{\theta}}_{k;t}] \notag
\end{align}
}
proving \eqref{E:rls-theta}.

Similarly obtained Formula for $\Lambda$:
{\allowdisplaybreaks
\begin{align}
    \Lambda_{k;t} & \leftarrow V_{k;y;t}+c_{l,k}y_{l;t}^2-\left(\bm{V}_{k;y\psi;t}+c_{l,k} \bm{\psi}_{l;t} y_{l;t}\right)\T \notag \\[1mm]
    {} &\times \left(\!\!\bm{C}_{k;t}\!
    -\! \frac{c_{l,k} \bm{C}_{k;t} \bm{\psi}_{l;t}\bm{\psi}_{l;t}\T \bm{C}_{k;t}}
     {1 + c_{l,k} \bm{\psi}_{l;t}\T \bm{C}_{k;t} \bm{\psi}_{l;t}}\!\right)\!\!
     \left( \bm{V}_{k;y\psi;t}+c_{l,k} \bm{\psi}_{l;t} y_{l;t}\right)\notag\\[1mm]
     {} &\leftarrow \Lambda_{k;t}+ \frac{\left(c_{l,k}y_{l;t}-c_{l,k}\bm{\psi}_{k;t}\T\widehat{\bm{\theta}}_{k;t}\right)^2}
             {1 + c_{l,k} \bm{\psi}_{l;t}\T \bm{C}_{k;t} \bm{\psi}_{l;t}}\notag
         \end{align}}
proves \eqref{E:rls-lambda}. Finally, the fact that
\begin{equation}
    \sum_{l\in\Nk} c_{l,k} = 1 \notag
\end{equation}
proves \eqref{E:rls-nu}.  \phantom{xxxxxxxxxxxxxxxxxxxxxxxxxxxxxxxxxx}\qed
\end{pf}
Obviously, since $c_{l,k}$ sum to unity, it is sufficient to increment $\nu_{k;t}$ at each time step by 1.

The well-known recursive least-squares evaluate a covariance matrix and the regression coefficients estimates, which is the same as $\bm{C}$ and $\widehat{\bm{\theta}}$ in the reparameterized $\NiG$ pdf. In this respect, the dynamic Bayesian diffusion estimation of the Bayesian regressive model is completely equivalent to the diffusion (unweighted) RLS, cf. \citet{Cattivelli2008Diffusion}. This proves the feasibility of the method. However, the exploited probabilistic framework allows to use the very general principles given in Section \ref{S:diffusion} with a wider class of various models.

\begin{algorithm}
    \DontPrintSemicolon
    \caption{Diffusion Bayesian regressive model}
    \textbf{Initialization:}\;
    \ForAll{$k\in\{1,\dots,M\}$}{
        Set prior statistics $\bm{V}_{k;0}$ and $\nu_{k;0}$. \;
        Set weights $c_{l,k}$ and $a_{l,k}$, $l\in\Nk$. \;
    }
    \BlankLine
    \textbf{Online steps:}\;
    \For{$t=1,2,\ldots$}{
        \textbf{Incremental update:}\;
        \ForAll{$k\in\{1,\dots,M\}$}{
            Gather data $[y_{l;t}, \bm{\psi}_{l;t}]\T$ for all $l\in\Nk$. \;
            Perform the updates of $\bm{V}_{k,t-1}, \nu_{k;t-1}$, Prop. \ref{P:gauss-incremental-update}. \;
            Calculate point estimates $\widehat{\bm{\theta}}_{k;t}$, Prop. \ref{T:bayesian-estimation-nig}. \;
        }
        \BlankLine
        \textbf{Spatial update:}\;
        \ForAll{$k\in\{1,\dots,M\}$}{
            Gather point estimates $\widehat{\bm{\theta}}_{l;t}$ for all $l\in\Nk$. \;
            Perform the update of $\widehat{\bm{\theta}}_{k;t}$, Prop. \ref{P:gauss-spatial-update}. \;
        }
    }
\end{algorithm}

\section{Conclusions}\label{S:conclusion}
The dynamic Bayesian diffusion estimation methodology provides a way to solving the decentralized estimation problems in the modern complex distributed systems, e.g., the sensor and ad-hoc networks. The theoretical aspects of the method are advocated by the maximum entropy and minimum cross-entropy principles. Being developed in the Bayesian framework, it is directly applicable to a wide class of different models. As a special case, the application of the methodology to the dynamic Bayesian linear regression yields particularly useful diffusion recursive least squares. This aspect also supports the assumption of validity of the method. In addition, it demonstrates that for practical purposes it is possible to leave the distribution-oriented perspective in favor of the traditional non-Bayesian reasoning.

The foreseen research activities comprise, among others, the analysis of properties of the diffusion estimator, the Bayesian estimation under specific constraints related, e.g., to bandwidth etc. Also, a probabilistic method for dynamic determination of the weighting coefficients $a_{l,k}$ and $c_{l,k}$ is of particular interest.

\section{appendix}
\begin{prop}[Sherman-Morrison formula]\label{A:sherman-morrison}{\ \\}
    Let $\bm{A} \in \mathbb{R}^{n\times n}$ be an invertible matrix and $\bm{u},\bm{v} \in \mathbb{R}^{n}$ two vectors. Then, the following equality holds,
    \begin{equation}
        \left( \bm{A} + \bm{u}\bm{v}\T \right)^{-1}
        = \bm{A}^{-1}
        - \frac{\bm{A}^{-1} \bm{u}\bm{v}\T \bm{A}^{-1}}{1 + \bm{v}\T \bm{A}^{-1} \bm{u}}.
        \notag
    \end{equation}
\end{prop}
\begin{pf}
    Trivial.\phantom{xxxxxxxxxxxxxxxxxxxxxxxxxxxxxxx}\qed
\end{pf}


\end{document}